# Time-Reversible, Symplectic, Angular Velocity Based Integrator for Rigid Linear Molecules

Somajit Dey
*Department of Physics, University of Calcutta, 92, A.P.C Road, Kolkata-700009*
E-mail: sdphys_rs@caluniv.ac.in

**Abstract**: A very simple explicit integrator for the rotational motion of rigid linear molecules is presented which can preserve the rigidity of the molecules without requiring any constraint force. The integrator is time-reversible and symplectic, thus preserving volume in phase space. It also conserves angular momentum. As expected, having all these virtues, it remains stable for large time-steps. Both the leap-frog and velocity-Verlet versions of the integrator are described. Since it features angular velocities explicitly, the integrator can be conveniently coupled to different thermostats. As a specific example, the Nosé-Hoover thermostatting is discussed in detail to aid ready implementation. A simpler and faster adaptation of the main integrator, appropriate for double precision computing, is also offered.

## I.  INTRODUCTION

The orientation of rigid *non-linear* molecules is usually represented by quaternions or rotation matrices [1, 2, 3, 4, 5]. For a *linear* molecule, however, the orientation can be simply described by a unit vector **e** along the molecular axis. Such a representation is also sufficient for any *axially symmetric* molecular model whose rotation about its axis is physically insignificant, e.g. the widely studied Gay-Berne spheroid [6]. The rotational velocity, conjugate to **e**, can be represented either by the angular velocity **ω** or the linear velocity **u** ($=\boldsymbol{\omega}\times\mathbf{e}$). A review of the standard integrators using both of these representations can be found in Ref. [7]. The symplectic constraint method LEN [7, 4], which uses the **e**, **u** representation, is considered the best among them as it allows the largest possible integration step-size. Consequently, in standard practice, LEN or some version of it [8] is usually adopted for simulations with rigid linear models. The present paper aims at highlighting a competitive alternative in the **e**, **ω** representation.

There exists a vast literature devoted to the integration of rotational motion of rigid bodies. In view of this, the present work may need some justification. In the works concerned with *general rigid bodies*, the main integration scheme is described in terms of quaternions, rotation matrices and angular momenta in body-fixed coordinates (since Euler's equations are formulated in body-fixed frame). *Linear or axially symmetric* molecules are treated as a special case in these articles, discussions on which, naturally, are still presented in terms of body-fixed coordinates and such [3, 5]. However, the field of molecular dynamics has seen (e.g. Ref. [6]) and continues to see the use of many axially symmetric models for which the abovementioned description involving quaternions, rotation matrices and body-fixed frames is completely unnecessary, as stated earlier. Actually, this exact simplicity in the dynamical description of axially symmetric models is one major reason for their popularity in modelling both small molecules (e.g. Ref. [9]) and parts of macromolecules and polymers (e.g. Ref. [10]). *To aid the studies of such models, therefore, a self-contained exposition of how to integrate the rotational motion of a linear rigid body in terms of its axial unit vector alone should be highly beneficial.* As yet, only Fincham has undertaken this important task by focussing on linear molecules proper [7]. The present paper now discusses a simple alternative to Fincham's methods. To the best of my knowledge, the present integrator did not appear explicitly in the literature or even the popular textbooks [11, 12, 13] on the subject of molecular dynamics. This is surprising given the simplicity, stability and hence applicability of this integrator, as will be evident from the following. In addition to the above, it may also be noted that the replacement of the obvious renormalisation scheme [Eq. (10)] with a computationally faster and easier feedback scheme [Eq. (12)], as described in Sec. III, is a novel strategy that is being put forward here for the first time.

## II.  EQUATIONS OF MOTION AND LEAP-FROG INTEGRATION

For linear molecules, the torque $\mathbf{T}$ and angular velocity $\boldsymbol{\omega}$ are perpendicular to $\mathbf{e}$. Denoting the moment of inertia as $I$, the equations of motion are

$$\frac{d\mathbf{e}}{dt} = \boldsymbol{\omega} \times \mathbf{e}, \tag{1}$$

$$\frac{d\boldsymbol{\omega}}{dt} = \frac{\mathbf{T}}{I}. \tag{2}$$

For clarity, in what follows, the dynamical quantities at *n*-th step will be denoted by suffix *n*.

Numerical integration of the above equations of motion can be done in direct analogy with the leap-frog algorithm for translational motion [12]. With states given by ($\boldsymbol{\omega}_{n-\frac{1}{2}}, \mathbf{e}_n$), the leap-frog scheme for rotational motion will thus entail an update of $\boldsymbol{\omega}$ using Eq. (2) with $\mathbf{e} = \mathbf{e}_n$ followed by an update of $\mathbf{e}$ using Eq. (1) with $\boldsymbol{\omega} = \boldsymbol{\omega}_{n+\frac{1}{2}}$. With a time-step $\tau$, the first update therefore is

$$\boldsymbol{\omega}_{n+\frac{1}{2}} = \boldsymbol{\omega}_{n-\frac{1}{2}} + \frac{\mathbf{T}_n}{I}\tau, \tag{3}$$

where $\mathbf{T}_n$ is the torque calculated with $\mathbf{e} = \mathbf{e}_n$. The on-step angular velocities required for kinetic energy computation can be interpolated from the mid-step ones as

$$\boldsymbol{\omega}_n = \frac{\boldsymbol{\omega}_{n+\frac{1}{2}} + \boldsymbol{\omega}_{n-\frac{1}{2}}}{2}. \tag{4}$$

Note that by virtue of $\mathbf{T}_n.\mathbf{e}_n = 0$, Eq. (3) and (4) together make

$$\boldsymbol{\omega}_{n+\frac{1}{2}}.\mathbf{e}_n = \boldsymbol{\omega}_{n-\frac{1}{2}}.\mathbf{e}_n = \boldsymbol{\omega}_n.\mathbf{e}_n. \tag{5}$$

Since $\boldsymbol{\omega}$ must be orthogonal to $\mathbf{e}$ at every time-step, i.e. $\boldsymbol{\omega}_n.\mathbf{e}_n = 0$, Eq. (5) demands that we choose initial conditions ($\boldsymbol{\omega}_{-\frac{1}{2}}, \mathbf{e}_0$) such that $\boldsymbol{\omega}_{-\frac{1}{2}}.\mathbf{e}_0 = 0$.

Now, the second leap-frog update amounts to a finite rotation of $\mathbf{e}$ about the direction of $\boldsymbol{\omega}_{n+\frac{1}{2}}$ by an angle $\omega_{n+\frac{1}{2}}\tau = 2\phi$, say. According to the rotation formula [14], this update becomes

$$\mathbf{e}_{n+1} = \mathbf{e}_n \cos 2\phi + (\frac{\boldsymbol{\omega}_{n+\frac{1}{2}}}{\omega_{n+\frac{1}{2}}} \times \mathbf{e}_n)\sin 2\phi + \frac{\boldsymbol{\omega}_{n+\frac{1}{2}}}{\omega_{n+\frac{1}{2}}^2}(\boldsymbol{\omega}_{n+\frac{1}{2}} \cdot \mathbf{e}_n)(1 - \cos 2\phi) \tag{6}$$

Inner product of $\boldsymbol{\omega}_{n+\frac{1}{2}}$ with both sides of Eq. (6) shows

$$\boldsymbol{\omega}_{n+\frac{1}{2}}.\mathbf{e}_{n+1} = \boldsymbol{\omega}_{n+\frac{1}{2}}.\mathbf{e}_n \tag{7}$$

which, by virtue of Eq. (5) and the initialisation $\boldsymbol{\omega}_{-\frac{1}{2}}.\mathbf{e}_0 = 0$, makes $\boldsymbol{\omega}_{n+\frac{1}{2}}.\mathbf{e}_n = 0$. Thus, the last term in Eq. (6) can be dropped. Truncation error analysis shows that Eq. (3) is third order accurate, viz. $\mathcal{O}(\tau^3)$ terms are neglected.

A similar third order approximation of Eq. (6) does away with the expensive computation of trigonometric functions but preserves the norm of $\mathbf{e}$ and hence the rigidity of the molecule. It uses the small angle approximations $\cos 2\phi = \frac{1-\phi^2}{1+\phi^2}$ and $\sin 2\phi = \frac{2\phi}{1+\phi^2}$, both of which are third order accurate as they neglect $\mathcal{O}(\phi^3)$. Eq. (6) thus turns into

$$\mathbf{e}_{n+1} = \frac{(1-\phi^2)\mathbf{e}_n + \tau\boldsymbol{\omega}_{n+\frac{1}{2}} \times \mathbf{e}_n}{1+\phi^2}. \tag{8}$$

Squaring both sides it is easily seen that $|\mathbf{e}_{n+1}| = |\mathbf{e}_n|$, i.e. norm is still preserved after the approximation.

Eq. (8) and (3) together constitute the present integrator. It must be noted that any initialisation for this leap-frog integrator must satisfy $|\mathbf{e}_0| = 1$ and $\boldsymbol{\omega}_{-\frac{1}{2}} \cdot \mathbf{e}_0 = 0$. Appendix B presents velocity-Verlet form of this integrator.

With its symmetry in the past and future coordinates, Eq. (3) is time-reversible. Such a symmetry is not readily apparent in Eq. (8). Hence, time-reversibility of Eq. (8) is clearly established in Appendix A. The integrator is therefore, time-reversible. The symplecticity of our integrator is addressed in Appendix B. Since both Eq. (3) and (8) respect the isotropy of space, the integrator conserves total angular momentum. Time-reversible, symplectic algorithms preserve volume in phase space and are known to possess good long-time stability [12, 15].

Since the angular velocity update in Eq. (3) is exactly similar in form to the familiar translational velocity update, various heat-baths can be coupled to $\boldsymbol{\omega}$ analogously. Nosé-Hoover thermostatting, for example, is discussed in Appendix C.

### III. ADAPTATIONS

Round-off errors will inevitably build up during finite-precision computation due to the preservation of norm by Eq. (8). As a result, norm of the on-step $\mathbf{e}$ continually drifts away from 1 thus compromising the rigidity of the molecules. To overcome this, Eq. (8) can be replaced by

$$\tilde{\mathbf{e}}_{n+1} = (1-\phi^2)\mathbf{e}_n + \tau\boldsymbol{\omega}_{n+\frac{1}{2}} \times \mathbf{e}_n, \tag{9}$$

$$\mathbf{e}_{n+1} = \frac{\tilde{\mathbf{e}}_{n+1}}{\sqrt{\tilde{\mathbf{e}}_{n+1} \cdot \tilde{\mathbf{e}}_{n+1}}}. \tag{10}$$

In *double precision computation*, however, stepwise round-off errors are smaller than typical truncation errors (which are third-order in time-step) by 7-8 orders of magnitude. Therefore, maintaining unit norm with the highest accuracy, such as that provided by Eq. (10), is not a strict necessity to achieve the desired accuracy of the thermodynamic and structural outputs. Therefore, if it helps reduce computational expenses, we can compromise a little in the preservation of norm. Motivated by this, let $\frac{(\tau\boldsymbol{\omega}_{n+\frac{1}{2}} \times \mathbf{e}_n)^2}{4}$ replace $\phi^2$ in Eq. (8) giving

$$\mathbf{e}_{n+1} = \frac{(1-\frac{\theta^2}{4})\mathbf{e}_n + \boldsymbol{\theta}}{1+\frac{\theta^2}{4}}, \tag{11}$$

where $\boldsymbol{\theta} = \tau\boldsymbol{\omega}_{n+\frac{1}{2}} \times \mathbf{e}_n$. This makes the algorithm more easy-to-implement since $\tau\boldsymbol{\omega}_{n+\frac{1}{2}} \times \mathbf{e}_n$ would have to be computed anyways. Additionally, it introduces a feedback mechanism that biases the norm towards unity at every step. This can be seen by squaring both sides of Eq. (11) which gives

$$\mathbf{e}_{n+1}^2 = \mathbf{e}_n^2 + \frac{\boldsymbol{\theta}^2}{(1+\frac{\theta^2}{4})^2}(1-\mathbf{e}_n^2). \tag{12}$$

This feedback (due to the second term in the right-hand side of Eq. (12)) checks long-term drift in norm but avoids the expensive square root computation of Eq. (10). Note that with infinite precision and $\mathbf{e}_0^2 = 1$, Eq. (12) would maintain exactly unit norm at each step. In that case, $\mathbf{\theta}^2$ would be equal to $\phi^2$ making Eq. (11) and (8) completely equivalent.

## IV. SIMULATION STUDIES

Working in double precision, molecular dynamics was performed on a system of 256 Gay-Berne (GB) molecules using both LEN and our method in its leap-frog version. Since we used double-precision, we used Eq. (11) when applying our method. Translational motion was solved by leap-frog too. Reduced units were used throughout. In order to have good energy conservation, the standard GB [6] was switched between radii $r_{cut-off}$ and $r_{switch}$ as $V(r) = s(r)V^{GB}(r)$, where

$$s(r) = 1 \text{ for } r \leq r_{switch}$$
$$= \frac{(r_{cut-off} + 2r - 3r_{switch})(r_{cut-off} - r)^2}{(r_{cut-off} - r_{switch})^3} \text{ for } r_{switch} < r \leq r_{cut-off} \quad (13)$$
$$= 0 \text{ for } r > r_{cut-off}$$

Parameterisation of Ref. [8] was followed. $r_{cut-off} = 3.9$ and $r_{switch} = 3.4$. The starting configuration was $\alpha$-FCC. Initial condition for a new production run was provided by the final configuration of a previous equilibration run of $2 \times 10^4$ steps where the desired energy was obtained by rescaling the velocities. Any instability was ascertained by drift or irregularity in energy or temperature [16]. Structural quantities were not calculated in stability studies since they are known to be somewhat less perceptive to instabilities [16]. Stabilities of LEN and our method were found to be almost the same in several runs with different moments of inertia, energies and time-steps. For example, with moment of inertia = 1, for energy near 5.5 (isotropic phase), both the integrators were unstable at $\tau = 0.004$ but well behaved at $\tau = 0.0035$. Fluctuations in energy and temperature in both cases were also nearly the same. Fig.1 compares the energy time-series of the competing integrators on the verge of instability ($\tau = 0.0038$). Fig.2 depicts the corresponding kinetic temperatures. We may observe that even in the vicinity of the region of instability, fluctuations in energies are only a few percent of those in the respective temperatures [16].

In regard to our method, the time-series of the thermodynamic quantities as computed by its different adaptations (Sec. III) consistently mirrored each other, definitely agreeing within the standard-deviation errors. This clearly suggests the promised viability of Eq. (11). A typical time-series showing the absence of long-time drift in norm is given in Fig.3.

## V. CONCLUSION

In summary, an explicit integrator for rigid linear molecules has been presented that was shown to be competitive with the highly stable constraint method LEN. Our integrator is based on angular velocity representation as opposed to the linear velocity representation of LEN. It owes its remarkable stability to its time-reversibility, symplecticity and preservation of volume in phase space. Some necessary easy-to-implement adaptations were also discussed.

The integrator discussed in this paper is readily applicable to molecular dynamics studies of uniaxial models such as rigid rods, spheroids, spherocylinders and linear poly-atoms (e.g. rigid dimers or trimers) with axially symmetric inter-molecular potentials (like Gay-Berne [6] or similar potentials [9]). This work concerned itself with integration of rotational motion of unconstrained linear units. A natural extension would be to ask how to integrate the constrained dynamics of flexible multibody chains composed of linked linear subunits. This will be dealt with in a forthcoming paper.

## ACKNOWLEDGEMENT

This work was done as part of a project sponsored by Council of Scientific & Industrial Research (CSIR), India under grant no. 09/028(0960)/2015-EMR-I.

# APPENDIX A:
# Time-reversibility of Eq. (8)

Taking the cross product of $\tau\boldsymbol{\omega}_{n+\frac{1}{2}}$ with both sides of Eq. (8) we get

$$\tau\boldsymbol{\omega}_{n+\frac{1}{2}} \times \mathbf{e}_{n+1} = \frac{(1-\phi^2)\tau\boldsymbol{\omega}_{n+\frac{1}{2}} \times \mathbf{e}_n - 4\phi^2 \mathbf{e}_n}{1+\phi^2}. \quad (A1)$$

Eliminating $\dfrac{\tau\boldsymbol{\omega}_{n+\frac{1}{2}} \times \mathbf{e}_n}{1+\phi^2}$ from Eq. (A1) and (8) we get

$$\tau\boldsymbol{\omega}_{n+\frac{1}{2}} \times \mathbf{e}_{n+1} = (1-\phi^2)(\mathbf{e}_{n+1} - \frac{(1-\phi^2)}{1+\phi^2}\mathbf{e}_n) - \frac{4\phi^2 \mathbf{e}_n}{1+\phi^2}, \quad (A2)$$

which, on simplification, produces

$$\mathbf{e}_n = \frac{(1-\phi^2)\mathbf{e}_{n+1} - \tau\boldsymbol{\omega}_{n+\frac{1}{2}} \times \mathbf{e}_{n+1}}{1+\phi^2}. \quad (A3)$$

Eq. (A3) is the same as Eq. (8) with $\mathbf{e}_n$ and $\mathbf{e}_{n+1}$ interchanged and $\boldsymbol{\omega}_{n+\frac{1}{2}}$ replaced by $-\boldsymbol{\omega}_{n+\frac{1}{2}}$. This demonstrates the time reversibility of Eq. (8). Reversibility can also be understood by noticing that Eq. (8) is equivalent to

$$\mathbf{e}_{n+1} = \mathbf{e}_n + \tau\boldsymbol{\omega}_{n+\frac{1}{2}} \times \mathbf{e}_{n+\frac{1}{2}} \quad (A4)$$

where $\mathbf{e}_{n+\frac{1}{2}} = \dfrac{(\mathbf{e}_{n+1} + \mathbf{e}_n)}{2}$. The explicit Eq. (8) can be arrived at from an implicit-looking Eq. (A4) by using the operator identity

$$(1 - \frac{\tau}{2}\boldsymbol{\omega}_{n+\frac{1}{2}}\times)^{-1} = \frac{1 + \frac{\tau}{2}(\boldsymbol{\omega}_{n+\frac{1}{2}}\times) + (\frac{\tau}{2})^2 \boldsymbol{\omega}_{n+\frac{1}{2}}(\boldsymbol{\omega}_{n+\frac{1}{2}}\cdot)}{1 + (\omega_{n+\frac{1}{2}}\frac{\tau}{2})^2}. \quad (A5)$$

# APPENDIX B:
# Velocity-Verlet form and Symplecticity

The algorithm implemented as velocity-Verlet is simply

$$\boldsymbol{\omega}_{n+\frac{1}{2}} = \boldsymbol{\omega}_n + \frac{\mathbf{T}_n}{I}\frac{\tau}{2} \quad (B1)$$

followed by Eq. (8) in one of its adaptations (Eq. (11) or Eq. (9)-(10)) and ultimately

$$\boldsymbol{\omega}_{n+1} = \boldsymbol{\omega}_{n+\frac{1}{2}} + \frac{\mathbf{T}_{n+1}}{I}\frac{\tau}{2}. \quad (B2)$$

It must be noted that any initialisation ($\boldsymbol{\omega}_0, \mathbf{e}_0$) for this velocity Verlet version must satisfy $|\mathbf{e}_0| = 1$ and $\boldsymbol{\omega}_0 \cdot \mathbf{e}_0 = 0$.

It can be easily shown that the angular velocity updates in Eq. (B1) and (B2) give the exact time evolution as generated by the kinetic part of the linear rigid body Hamiltonian. Eq. (8), on the other hand, gives the exact propagation due to the potential part of that Hamiltonian. Each of these propagations satisfies the symplectic condition individually. Since our integrator above is given as the sequence of these symplectic updates (viz. Eq. (B1) followed by Eq. (8) and (B2)), it must be symplectic itself [12].

## APPENDIX C:
## Coupling to Nosé-Hoover (NH) thermostat

Equations of motion for the NH canonical dynamics [17, 18, 19] are

$$\frac{d\mathbf{r}}{dt} = \frac{\mathbf{p}}{m}, \tag{C1}$$

$$\frac{d\mathbf{p}}{dt} = \mathbf{F} - v_1 \zeta^1 \mathbf{p}, \tag{C2}$$

$$\frac{d\boldsymbol{\omega}}{dt} = \frac{\mathbf{T}}{I} - v_2 \zeta^2 \boldsymbol{\omega}, \tag{C3}$$

$$\frac{d\zeta^i}{dt} = v_i (\frac{\Gamma^i}{\Theta} - 1), \tag{C4}$$

supplemented by Eq. (1). Above $\mathbf{r}, \mathbf{p}, \mathbf{F}$ denote position, momentum, and force. $\zeta^1$ and $\zeta^2$ denote the heat flow variables of the extended system corresponding to translation and rotational motion respectively. $v_1$ and $v_2$ are the coupling rates of the translational and rotational degrees of freedom to the heat bath. These coupling rates may be made equal to the characteristic frequencies (vibrational, librational, rotational or collisional) of the system, presumably extracted from sample runs, as is seen fit [19]. $\Theta$ above denotes the desired temperature and $\Gamma^i$ the kinetic temperature for translation ($i=1$) or rotation ($i=2$). The advantage of NH dynamics over constraint methods like that of Evans and Morriss is that Eq. (C1)-(C4) along with Eq. (1) satisfy the Gibbsian canonical distribution. This can easily be shown for rotational motion following the methods for pure translational motion [17, 20] with the proviso $\mathbf{T} = \nabla_{\mathbf{e}} V(\mathbf{r},\mathbf{e}) \times \mathbf{e}$. Observing the similarity of Eq. (C3) to (C2) it is integrated analogously [19] as

$$\boldsymbol{\omega}_{n+\frac{1}{2}} = \frac{\boldsymbol{\omega}_{n-\frac{1}{2}}(1 - v_2 \zeta_n^2 \frac{\tau}{2}) + \frac{\mathbf{T}_n}{I}\tau}{(1 + v_2 \zeta_n^2 \frac{\tau}{2})} \tag{C5}$$

with

$$\zeta_n^i = \zeta_{n-1}^i + v_i (\frac{\Gamma_{n-\frac{1}{2}}^i}{\Theta} - 1). \tag{C6}$$

Coupled with Eq. (8) in one of its adaptations (Eq. (11) or Eq. (9)-(10)), this completes the update of rotational degrees of freedom. Accuracy of the algorithm can be checked by the conservation of

$$E_n + \sum_{i=1,2} \left( \frac{1}{2} (\xi_n^i)^2 g_i k\Theta + v_i g_i k\Theta \int_0^{n\tau} \xi^i dt \right) \tag{C7}$$

where $k$ is the Boltzmann constant, $E_n$ is the on-step mechanical energy and $g_i$ denotes the number of translational degrees of freedom for $i=1$ and rotational degrees of freedom for $i=2$ [18]. The integral in Eq. (C7) can be calculated as

$$\int_0^{n\tau} \xi^i dt = \left( \sum_{j=1}^{n-1} \xi_j^i + \frac{\xi_0^i + \xi_n^i}{2} \right) \tau. \tag{C8}$$

It may be noted that the thermostatting above can also be extended to NPT dynamics [17, 20].

# FIGURES

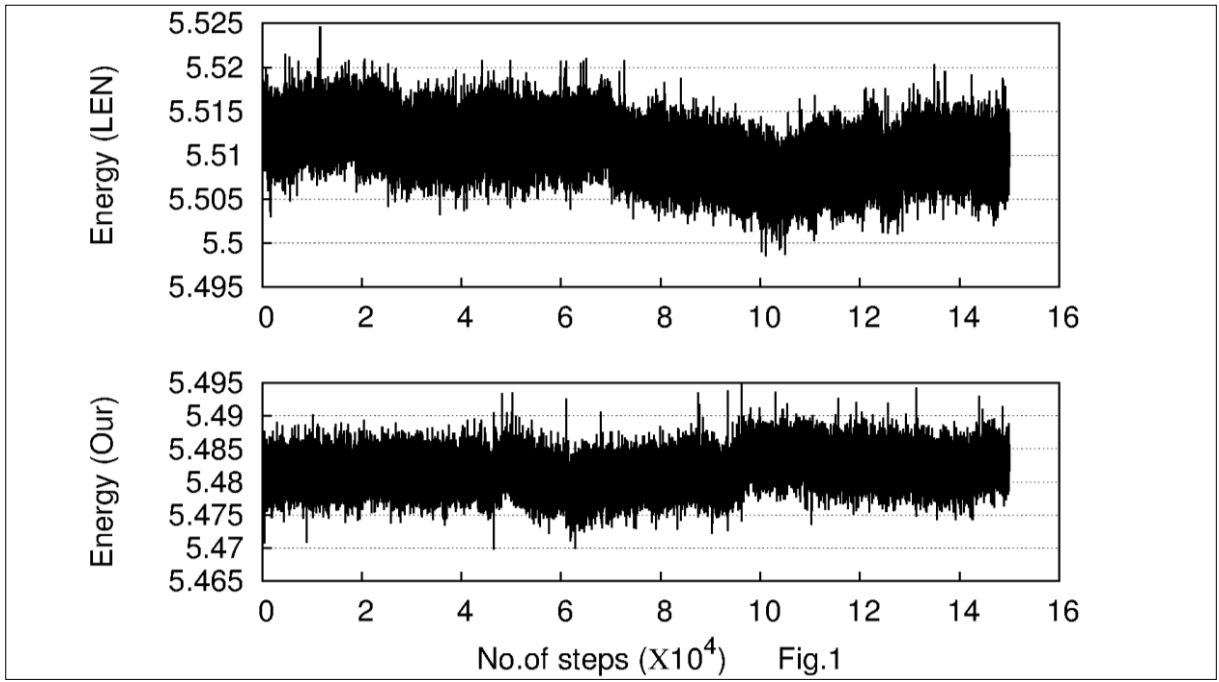
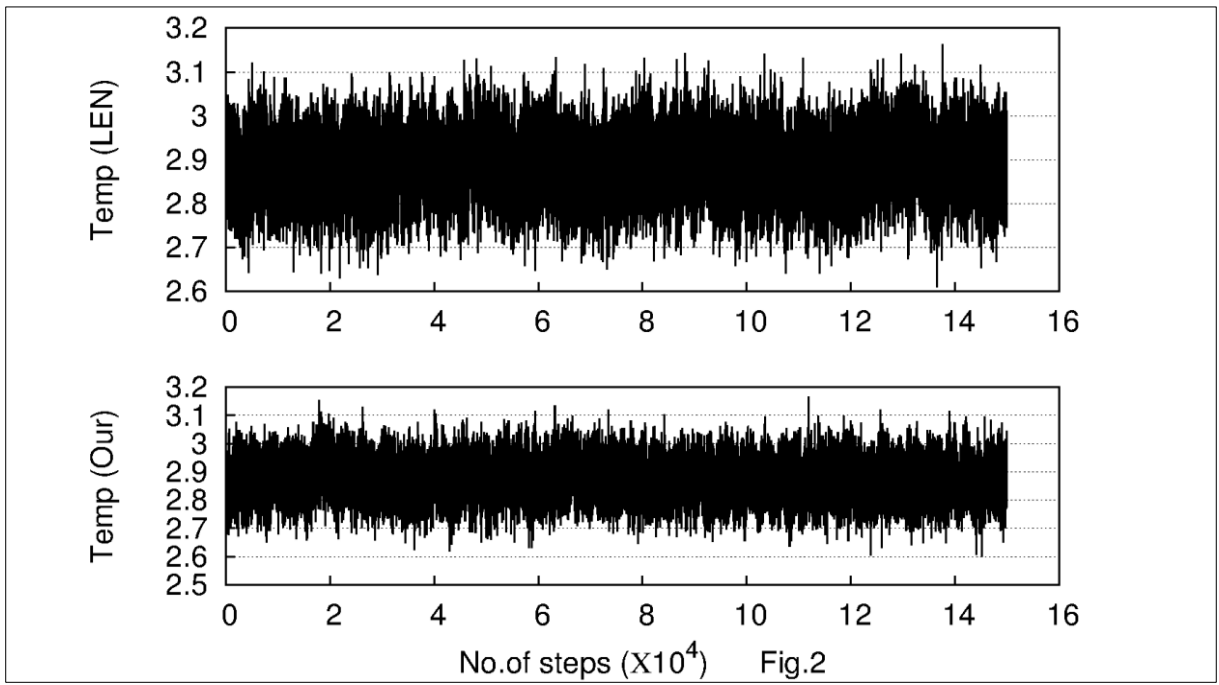

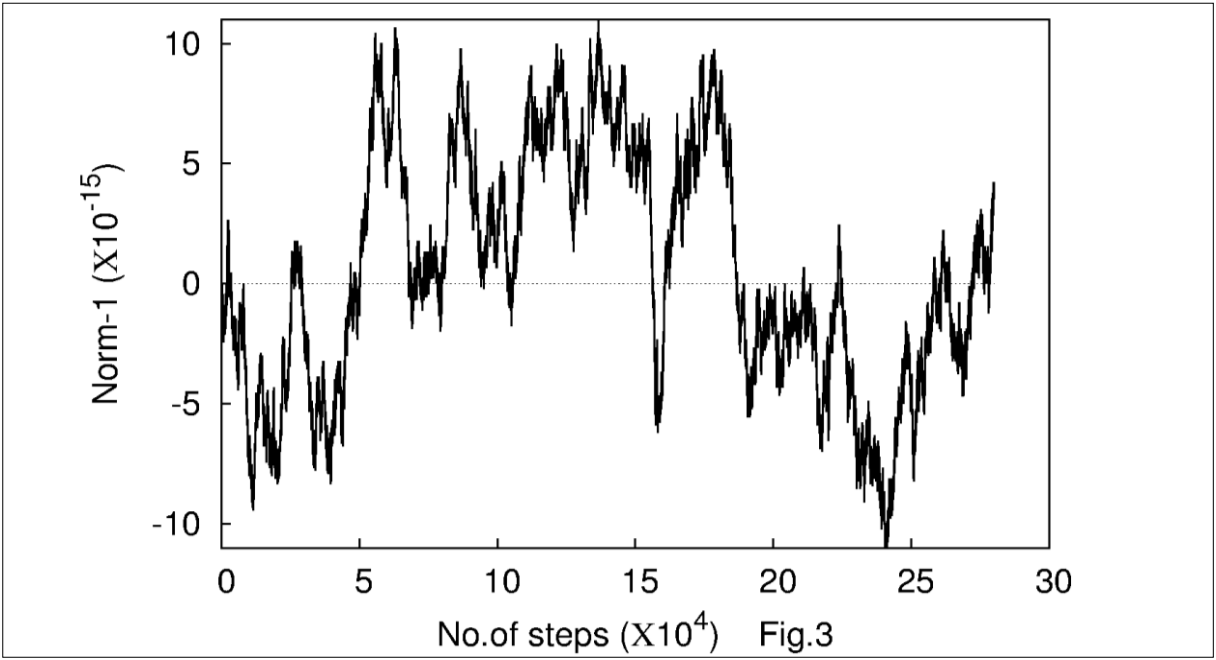
Fig.3